\documentclass[12pt]{amsart}

\usepackage{amsmath, amsfonts,amsthm,times,graphics}

\usepackage[active]{srcltx}

 \makeatletter
\renewcommand*\subjclass[2][2010]{%
  \def\@subjclass{#2}%
  \@ifundefined{subjclassname@#1}{%
    \ClassWarning{\@classname}{Unknown edition (#1) of Mathematics
      Subject Classification; using '2010'.}%
  }{%
    \@xp\let\@xp\subjclassname\csname subjclassname@#1\endcsname
  }%
}

 \makeatother
\newtheorem{theorem}{Theorem}[section]

\newtheorem{corollary}[theorem]{Corollary}

\newtheorem{proposition}[theorem]{Proposition}

\theoremstyle{definition}
\newtheorem{definition}[theorem]{Definition}

 \newtheorem{example}[theorem]{Example}

\newtheorem{remark}[theorem]{Remark}

 \makeatletter
\renewcommand*\subjclass[2][2010]{%
  \def\@subjclass{#2}%
  \@ifundefined{subjclassname@#1}{%
    \ClassWarning{\@classname}{Unknown edition (#1) of Mathematics
      Subject Classification; using '1991'.}%
  }{%
    \@xp\let\@xp\subjclassname\csname subjclassname@#1\endcsname
  }%
}

 \makeatother

\begin{document}
\title[A generalization of some random variables
 \ldots]{A generalization of some random variables 
involving in certain compressive sensing problems}
\author{Romeo Me\v strovi\' c}
\address{Maritime Faculty Kotor, University of Montenegro, 
85330 Kotor, Montenegro} 
\email{romeo@ac.me}

 
\begin{abstract}
In this paper we give a generalization of the discrete complex-valued 
random variable defined and investigated  in \cite{ssa} and \cite{m8}. 
We prove the statements concerning the expressions for the excepted value and the 
variance of this random variable.

In partucular, such a random variable here is  defined for each of $m$ rows 
of any $m\times N$ complex or real matrix 
${\rm{\bf A}}$ with $1\le m\le N$. We  consider
 the arithmetic mean $\bar{X}(m)$ of these $m$ random variables
and we   deduce the expressions for the  expected value
$\Bbb E[\bar{X}(m)]$  and the variance 
${\rm{ Var}}[\bar{X}(m)]$ 
 of $\bar{X}(m)$.
Using the expression for ${\rm{ Var}}[\bar{X}(m)]$, 
we establish some equalities and inequalities 
involving ${\rm{Var}}[\bar{X}(m)]$, 
the Frobenius norm, the largest eigenvalue,  
the largest singular value and the coherence of a matrix ${\rm{\bf A}}$. 
 It is showed that some of  these estimates are 
 closely related to the Welch bound of the coherence of a 
$m\times N$ complex or real matrix 
${\rm{\bf A}}$ with  $1\le m\le N$.
Taking into account that the value of  coherence of the 
measurement matrix in the  theory of compressive sensing
has a significant role, we believe that our results should be useful for 
some topics of this theory. 
  \end{abstract}  
    \maketitle

\section{Introduction and Preliminaries}

As usually, throughout our considerations we use the 
term ``multiset'' (often written as ``set'') to mean ``a totality having possible 
multiplicities''; so that two (multi)sets will be counted as equal if 
and only if they have the same elements with identical multiplicities.
Let $\Bbb C$ and $\Bbb R$ denote the fields of complex and real 
numbers, respectively. For a given positive integer $N$, let ${\mathcal M}_N$ 
denote the collection of all multisets of the form 
   $$
\Phi_N=\{z_1,z_2,\ldots ,z_N\},\leqno (1)
   $$
where $z_1,z_2,\ldots ,z_N\in\Bbb C$ are arbitrary (not necessarily 
distinct) complex numbers.
Furthermore, denote by  ${\mathcal M}$  the set consisting 
of all multisets of the form (1), i.e.,  
   $$
{\mathcal M}=\bigcup_{N=1}^{\infty}{\mathcal M}_N. 
  $$

Following Definition 1.2 from \cite{m8} 
(also see \cite[Section 2]{ssa},
\cite[Section II]{sso},  \cite[Definition 1.1]{m7}
and  \cite[Definition 1.1]{m7}), the random 
variable $X(m, \Phi_N)$ can be generalized  as follows.
   
  \begin{definition}
Let $N$ and $m$  be arbitrary nonnegative integers 
such that  $1\le m\le N$.
For given not necessarily distinct complex numbers
$z_1,z_2,\ldots ,z_N$, let  
$\Phi_N \in {\mathcal M}_N$ be a multiset defined by (1).
 Define the discrete complex-valued random variable $X(m,\Phi_N)$ as
  \begin{eqnarray*}
 && \mathrm{Prob}\left(X(m,\Phi_N)
 = \sum_{i=1}^m z_{n_i}\right)\\
(2)\qquad\qquad &= &\frac{1}{{N\choose m}}\cdot \big|\{\{t_1,t_2,\ldots,t_m\}
\subset\{1,2,\ldots,N\}: 
\sum_{i=1}^m z_{t_i}=\sum_{i=1}^mz_{n_i} \big\}|
\quad\qquad\\
& =&:\frac{q(n_1,n_2,\ldots, n_m)}{{N\choose m}},
  \end{eqnarray*}
  where $\{n_1,n_2,\ldots, n_m\}$ is an arbitrary fixed
 subset of $\{1,2,\ldots,N\}$ such that $1\le n_1<n_2<\cdots <n_m\le N$;
moreover, $q(n_1,n_2,\ldots, n_m)$ is the cardinality of a collection
 of all subsets
$\{t_1,t_2,\ldots,t_m\}$ of the set $\{1,2,\ldots,N\}$ such that 
$\sum_{i=1}^m z_{t_i}=\sum_{i=1}^mz_{n_i}$.
  \end{definition}

Notice that the above definition is correct  taking into account that there 
are ${N\choose m}$ index sets $T\subset \{1,2,\ldots,N \}$ with 
$m$ elements. Moreover, a very  short, but not 
strongly exact version of Definition 1.1  is given as follows 
(cf. \cite[Definition 1.2']{m8}).
   \vspace{2mm}

  \noindent {\bf Definition 1.1'.}
Let $N$ and $m$  be arbitrary nonnegative integers 
such that  $1\le m\le N$.
For given not necessarily distinct complex numbers
$z_1,z_2,\ldots ,z_N$, let  
$\Phi_N \in {\mathcal M}_N$ be a multiset defined by (1).
Choose a random subset  $S$ of size $m$ (the so-called $m$-element 
subset) without replacement from the set 
$\{1,2,\ldots, N\}$. Then the 
 complex-valued discrete random variable $X(m,\Phi_N)$ is defined as a sum
  $$
X(m,\Phi_N)=\sum_{n\in S}z_n.
  $$
  \vspace{1mm}

Accordingly, here we prove the following result 
(cf. proof of Theorem 2.4 in \cite{m8} as a particular case).

 \begin{theorem} 
Let $N$ and $m$ be positive integers such that
  $N\ge 2$ and  $1\le m\le N$. Let $\Phi_N=\{z_1,z_2,\ldots ,z_N\}$ be any 
multiset with $z_1,z_2,\ldots ,z_N\in \Bbb C$. 
 Then the expected value and 
the variance  of the random variable $X(m,\Phi_N)$  
from Definition $1.1$  are 
respectively given  by
  $$
\Bbb E[X(m,\Phi_N)]=\frac{m}{N}\sum_{i=1}^Nz_i\leqno(3) 
   $$
and
    $$
   {\rm Var}[X(m,\Phi_N)]=
\frac{m(N-m)}{N^2(N-1)}\left(N \sum_{i=1}^N |z_i|^2 - 
 \big| \sum_{i=1}^N z_i \big|^2\right).\leqno(4)
    $$
Moreover, the second moment of the random variable $|X(m,\Phi_N)|$  
is given  by
      $$
   \Bbb E[|X(m,\Phi_N)|^2]=
\frac{m}{N(N-1)}\left((N-m) \sum_{i=1}^N |z_i|^2 + 
 (m-1)\big| \sum_{i=1}^N z_i \big|^2\right).\leqno(5)
    $$
 \end{theorem}
As a particular case    of Theorem 1.2, we immediately obtain 
the following straightforward result.

\begin{corollary}
Under notations and the  assumptions of Theorem $1.2$, for $m=N$ we have
 $$
\Bbb E[X(N,\Phi_N)]=\sum_{i=1}^Nz_i, 
   $$
    $$
   {\rm Var}[X(N,\Phi_N)]=0
    $$
and
      $$
   \Bbb E[|X(N,\Phi_N)|^2]=
\big| \sum_{i=1}^N z_i \big|^2.
    $$
 \end{corollary}  
 
Another consequence of Theorem 1.2 is given as follows.

\begin{corollary}
Under notations and the  assumptions of Theorem $1.2$, we have
  $$ 
{\rm Var}[X(m,\Phi_N)]=
\frac{m(N-m)}{N^2(N-1)}\sum_{1\le i<k\le N}|z_i-z_k|^2.\leqno(6)  
   $$
\end{corollary}

Moreover, for our purposes concerning the Welch bound given in the 
next setion, we will need the following immediate consequence of 
Theorem 1.2.

\begin{corollary}
Under notations and the  assumptions of Theorem $1.2$, 
let $\bar{X}(m,\Phi_N)$ be   the random variable defined as
  $$
\bar{X}(m,\Phi_N)=\frac{X(m,\Phi_N)}{m}.
  $$ 
Then 
$$
\Bbb E[\bar{ X}(m,\Phi_N)]=\frac{\sum_{i=1}^Nz_i}{N}\leqno(7) 
   $$
and
    $$
   {\rm Var}[\bar{X}(m,\Phi_N)]=
\frac{N-m}{mN^2(N-1)}\left(N \sum_{i=1}^N |z_i|^2 - 
 \big| \sum_{i=1}^N z_i \big|^2\right).\leqno(8)
  $$
   \end{corollary}
\begin{remark} Recall that  the random variable $\bar{X}(m,\Phi_N)$ 
 from Corollary 1.5 can be considered as the ``random arithmetic mean''
of $m$ randomly chosen elements of  a set $\Phi_N=\{z_1,z_2,\ldots ,z_N\}$.
Notice also  that a two-dimensional analogue  of  this random variable 
is investigated in \cite{m5}.
   \end{remark}

\begin{example}
Quite recently,  the author of this paper \cite{m8}
(also see \cite{m7} and \cite{m3}) 
considered some complex-valued discrete random 
variables  $X_l(m,N)$ ($0\le l\le N-1$, 
$1\le M\le N$), which are closely related 
to the random variables investigated by LJ. Stankovi\'c, S. Stankovi\'c and 
M. Amin in \cite{ssa}. Note that in view of \cite[Definition 1.2]{m8}, such a 
random variable $X_l(m,N)$ ($0\le l\le N-1$, $1\le M\le N$) is a particular case 
of the random variable $X(m,\Phi_N)$ from Definition 1.1 associated 
to the multiset $\Phi_N=\Phi(l,N):=\{e^{-j2nl\pi/N}:\, n=1,2,\ldots,N\}$
($j$ is the imaginary unit). 
If $l\not= 0$, then $\sum_{n=1}^Ne^{-j2nl\pi/N}=0$. Taking this together 
with $|z_n|=|e^{-j2nl\pi/N}|=1$ ($n=1,2,\ldots,N$) into the expressions 
(3), (4) and (5) of Theorem 1.2, we immediately obtain the 
equalities (18) and (19) from \cite[Theorem 2.4]{m8} as follows.
 $$
\Bbb E[X_l(m,N)]=0\leqno(9) 
   $$
and
   $$
{\rm Var}[X_l(m,N)]=\Bbb E[|X_l(m)|^2]=\frac{m(N-m)}{N-1}.\leqno(10)
   $$
\end{example}

The remainder of the paper is organized as follows.
According to  Definition 1.1, for a complex or real matrix
${\rm{\bf A}}$ with $m$ rows and $N$ columns ($m\le N$),
we define the discrete complex-valued random variable $X_i(m)$
associated to the $i$th row of ${\rm{\bf A}}$, $i=1,2,\ldots, m$ (considered
 as a multiset  $\Phi_N$ from Definition 1.1).
Next we consider the arithmetic mean of these $m$ random variables
and deduce the expresions for its expected value and the variance.
Using these expressions, we establish some equalities and estimates 
involving the Frobenius norm, the largest eigenvalue,  
the largest singular value and the coherence of a matrix ${\rm{\bf A}}$.
Notice that in  some of these equalities and  estimates
the  Welch bound for coherence of a  matrix is involved.
We also point out that the value of  coherence of  the 
measurement matrix in the  theory of compressive sensing
has a significant role. Proofs of the results are given in Section 3.

\section{A probabilistic approach to the Welch bound}

We start with the  basic definition and results concerning the coherence of a 
matrix and its lower bound.  Here, as always in the sequel,
${\Bbb K}^{m\times N}$ denotes the space of all matrices over the field
  $\Bbb K$ with $m$ rows and $N$ columns. 

Let  ${\rm{\bf A}}=[a_{ik}]_{m\times N}\in {\Bbb K}^{m\times N}$ be a matrix 
with $l_2$-normalized columns 
${\rm{\bf a_1}},{\rm{\bf a_2}},\ldots , {\rm{\bf a_N}}$, 
where the field $\Bbb K$ can either be $\Bbb R$ or $\Bbb C$.
This means that 
    $$
\Vert {\rm{\bf a_k}}\Vert_2:=\sqrt{\sum_{i=1}^m |a_{ik}|^2}=1\quad
{\rm \,\, for\,\, all}\quad k=1,2,\ldots, N.
     $$
The {\it coherence} $\mu:=\mu({\rm{\bf A}})$ of the matrix ${\rm{\bf A}}$ 
is defined as 
    $$
\mu:=\max_{1\le i\not= k\le N} |\langle {\rm{\bf a_i}},{\rm{\bf a_k}}\rangle|,
    $$ 
where $\langle \cdot, \cdot  \rangle$ is the scalar product in the vector 
space  $\Bbb K^m$. 

The significance of the value of  coherence of the 
measurement matrix in the {\it theory of compressive sensing}
was shortly explained in \cite[Chapter 5, p. 111]{fr} as: 
``{\it In compressive sensing, the analysis of recovery algorithms usually 
involves a quantity that measures the suitability of 
the  measurement matrix. The coherence is a very simple such measure of
quality. In general, the smaller the coherence, the better the recovery 
algorithms perform.}''  For more 
information on the development of compressive sensing 
(also known as {\it compressed sensing}, 
{\it compressive sampling}, or {\it sparse recovery}), 
see \cite{do}, \cite{fr}, \cite[Chapter 10]{s1}, \cite{sdt}, \cite{op}
and \cite{sos}.

It was noticed in \cite[p. 159]{ss} (also see \cite{s2} and 
\cite[Remark 2.11]{m8}) that the ratio 
$\sqrt{\frac{N-m}{m(N-1)}}$ ($1\le m\le N$) is closely related to the variance 
of the random variable from Example 1.7 givan by (10), i.e.,
  $$
\sqrt{\frac{N-m}{m(N-1)}}=\frac{\sigma[X_l(m,N)]}{m}:=
\frac{\sqrt{{\rm Var}[X(m,\Phi_N)]}}{m}.\leqno(11)
   $$  
 This ratio  is a crucial parameter 
(known as the  {\it Welch bound} \cite{we}) for coherence 
$\mu$ of a  measurement matrix ${\rm{\bf A}}$ for corrected signal detection.
More precisely  (for a particularly 
elegant and very short proof of this bound see \cite{jmf}; also see 
\cite[Chapter 5, Theorem 5.7]{fr}), 
the coherence $\mu$ of a matrix ${\rm{\bf A}}\in {\Bbb K}^{m\times N}$, 
where the field $\Bbb K$ can either be $\Bbb R$ or $\Bbb C$, 
with $l_2$-normalized columns satisfies the inequality
   $$
\mu\ge \sqrt{\frac{N-m}{m(N-1)}},\leqno(12)
  $$  
which under  notation of Example 1.7  can be written as 
   $$
\mu \ge \frac{\sigma[X_l(m,N)]}{m}.\leqno(13)
   $$
Equality in the above two inequalities holds if and only if the columns 
$\rm{\bf{a}}_1,\ldots ,\rm{\bf{a}}_N$ of the matrix ${\rm{\bf  A}}$
form an {\it equiangular tight frame} 
(see \cite[Chapter 5, Definition 5.6]{fr}). Ideally,
in the theory of compressive sensing the coherence $\mu$
of a measurement matrix   ${\rm{\bf A}}$ should be small (see 
\cite[Chapter 5]{fr},  \cite{cp}, \cite{det} and \cite{sh}). 
 Let us observe  that if $m \ll N$, then this bound reduces to 
approximately $\mu\ge 1/\sqrt{m}$. There is a lot of possible ways 
to construct matrices with small coherence. Not surprisingly, 
one possible option is to consider random matrices ${\rm{\bf A}}$ with 
each entry generated independently at random 
(cf. \cite[Chapter 11]{pi}).

The above observations and the expression (4) of  Theorem 1.2
give the motivation for obtaining a  probabilistic approach 
to the Welch bound described as follows. 
Let  ${\rm{\bf A}}=[a_{ik}]_{m\times N}\in {\Bbb K}^{m\times N}$ be any matrix
 ($\Bbb K=\Bbb R$ or $\Bbb K=\Bbb C$). The 
{\it Frobenius norm} $\Vert {\rm{\bf A}}\Vert_F$, 
sometimes also called the {\it Euclidean norm}, is a matrix norm of an $m\times N$ matrix
 ${\rm{\bf A}}=[a_{ik}]_{m\times N}$ defined as (see, e.g., \cite{gv} or 
\cite[p. 524]{fr}) 
   $$
\Vert {\rm{\bf A}}\Vert_F= \sqrt{\sum_{i=1}^m\sum_{k=1}^N|a_{ik}|^2}.
\leqno(14)
   $$
It is also equal to the square root of the {\it matrix trace} of
${\rm{\bf A}}{\rm{\bf A}}^*$, where ${\rm{\bf A}}^*$ is the 
{\it conjugate transpose}, i.e., 
  $$
\Vert {\rm{\bf A}}\Vert_F=
\sqrt{{\rm tr}({\rm{\bf A}}{\rm{\bf A}}^*)}.\leqno(15)
  $$   
Denote by $\Bbb K^{m\times N}$ the set of 
all matrices over the field $\Bbb K$ with $m$ rows and $N$ columns. 
Notice that after identifying matrices on $\Bbb K^{m\times N}$ with 
vectors in $\Bbb K^{mn}$, the Frobenius norm can be interpreted as an 
$l_2$-norm on $\Bbb K^{mn}$.    

Recall also that the 
{\it spectral norm} 
(or often called the {\it operator norm on} $l_2$) 
$\Vert {\rm{\bf A}}\Vert_2$
of the matrix ${\rm{\bf A}}\in {\Bbb K}^{m\times N}$ is defined as
   $$
\Vert {\rm{\bf A}}\Vert_2=
\sup\left\{\frac{\Vert {\rm{\bf Ax}}\Vert_2}{\Vert x\Vert_2}: 
\, x\in \Bbb K^N\quad {\rm with}\quad   x\not= 0\right\}.\leqno(16)
  $$    
It is known that (see, e.g., 
\cite[Lemma A.5, p. 519 and (A.17), p. 524]{fr})
  $$
 \Vert {\rm{\bf A}}\Vert_2
=\sqrt{\lambda_{\max}({\rm{\bf A}}^*{\rm{\bf A}})}=
 \sigma_{\max}({\rm{\bf A}})\le 
\sqrt{\sum_{i=1}^m\sum_{k=1}^N|a_{ik}|^2}=\Vert {\rm{\bf A}}\Vert_F,\leqno(17) 
  $$
where $\lambda_{\max}({\rm{\bf A}}^*{\rm{\bf A}}^*)$
is the largest {\it eigenvalue} of the matrix ${\rm{\bf A}}^*{\rm{\bf A}}$ 
and  $\sigma_{\max}({\rm{\bf A}})$ is  the largest {\it singular value} 
of the  matrix ${\rm{\bf A}}$. The equality in the inequality 
from (17) holds if and only if the matrix
${\rm{\bf A}}$ is a rank-one matrix or a zero matrix. Furthermore,
 it is well known that (see, e.g., \cite[Remark A.6 (a), p. 521]{fr})  
  $\Vert {\rm{\bf A}}\Vert_2=\Vert {\rm{\bf A}}\Vert_2^*$.

For the  purpose of this section, we give the following definition.

\begin{definition} Let 
${\rm{\bf A}}=[a_{ik}]_{m\times N}\in {\Bbb K}^{m\times N}$
($\Bbb K=\Bbb R$ or  $\Bbb K=\Bbb C$) be a
matrix with $1\le m\le N$. For any fixed $i=1,2,\ldots , m$ 
put  $A_i=\{a_{i1},a_{i2},\ldots , a_{iN} \}$. 
We say that $A_i$ is the multiset associated to the $i$th row
${\rm{\bf A}}_i:=(a_{i1},a_{i2},\ldots , a_{iN})$ of the matrix 
${\rm{\bf A}}$. Then by Definition 1.1, for each $i=1,2,\ldots,m$  we  
define the discrete complex-valued random variable $X(m,A_i):=X_i(m)$
which corresponds to the multiset  $\Phi_N:=A_i$ from Definition 1.1.
  \end{definition}
The following statement can be easily deduced from Theorem 1.2.

\begin{proposition} Let 
${\rm{\bf A}}=[a_{ik}]_{m\times N}\in {\Bbb K}^{m\times N}$
$(\Bbb K=\Bbb R$ or  $\Bbb K=\Bbb C)$ be a
matrix with $1\le m\le N$. Define the discrete 
complex-valued random variable $\bar{X}(m,{\rm{\bf A}}):=\bar{X}(m)$
as 
  $$
\bar{X}(m)=\frac{\sum_{i=1}^m X_i(m)}{m},\leqno(18)
  $$
where $X_i(m)$ $(i=1,2,\ldots,m)$ are random variables defined by Definition
$2.1$.  Then the expected value and
the variance of the random variable $\bar{X}(m)$ 
are respectively given by 
  $$
\Bbb E[\bar{X}(m)]=\frac{1}{N}\sum_{i=1}^m\sum_{k=1}^Na_{ik}\leqno(19)
  $$ 
and
 $$
{\rm Var}[\bar{X}(m)]=\frac{N-m}{mN^2(N-1)}
\left(N \sum_{i=1}^m\sum_{k=1}^N|a_{ik}|^2-
\sum_{i=1}^m|\sum_{k=1}^Na_{ik}|^2\right),\leqno(20)
 $$
or equivalently,
  $$
{\rm Var}[\bar{X}(m)]=\frac{N-m}{mN^2(N-1)}
\left(N \Vert {\rm{\bf A}}\Vert_F^2 -
\sum_{i=1}^m|\sum_{k=1}^Na_{ik}|^2\right),\leqno(21)
 $$
where  $\Vert {\rm{\bf A}}\Vert_F$ is the  Frobenius norm of the matrix
${\rm{\bf A}}$  defined by $(14)$.
\end{proposition}

Notice that applying the expressions (19) and (20) of Proposition 2.2,
in \cite{m6} the author of this paper deduced   some combinatorial 
identities. As applications, several  variations and generalizations  
of remarkable {\it Chu-Vandermonde identity} are established in \cite{m6}.

Taking the inequality from (17) into  the expression  (21) of Proposition 
2.2, immediately gives the following result.

\begin{corollary}
Let ${\rm{\bf A}}=[a_{ik}]_{m\times N}\in 
{\Bbb K}^{m\times N}$
$(\Bbb K=\Bbb R$ or  $\Bbb K=\Bbb C)$ be a
matrix with $1\le m\le N$. Then under notations of Proposition $2.2$, 
the following inequality holds:
 $$
{\rm Var}[\bar{X}(m)]\ge \frac{N-m}{mN^2(N-1)}
\left(N\sigma_{\max}^2({\rm{\bf A}}) -
\sum_{i=1}^m|\sum_{k=1}^Na_{ik}|^2\right),\leqno(22)
 $$
where $\sigma_{\max}({\rm{\bf A}})$ is  the largest singular value 
of the matrix ${\rm{\bf A}}$$;$ or equivalently,
  $$
{\rm Var}[\bar{X}(m)]\ge \frac{N-m}{mN^2(N-1)}
\left(N\lambda_{\max}^2({\rm{\bf A}}^*{\rm{\bf A}}) -
\sum_{i=1}^m|\sum_{k=1}^Na_{ik}|^2\right),\leqno(23)
 $$
where  $\lambda_{\max}({\rm{\bf A}}^*{\rm{\bf A}}^*)$
is the largest eigenvalue of the matrix ${\rm{\bf A}}^*{\rm{\bf A}}$.
\end{corollary}

Moreover, the equality (21) of Proposition 2.2 directly implies
the following inequality.
    
\begin{corollary}
Let ${\rm{\bf A}}=[a_{ik}]_{m\times N}\in 
{\Bbb K}^{m\times N}$
$(\Bbb K=\Bbb R$ or  $\Bbb K=\Bbb C)$ be a
matrix with $1\le m\le N$. Then under notations of Proposition $2.2$, 
the following inequality holds:
   $$
{\rm Var}[\bar{X}(m)]\le\frac{N-m}{mN(N-1)}
\Vert {\rm{\bf A}}\Vert_F^2,
\leqno(24)
 $$
where equality holds if and only if $\sum_{k=1}^Na_{ik}=0$ 
for all $i=1,2,\ldots,m$ $($i.e., if and only if the sum of all 
entries in every row of the matrix   ${\rm{\bf A}}$ is equal to 
zero$)$.
\end{corollary}

As an immediate consequence of Proposition 2.2 and the relations 
of (17), we also  obtain the following assertion.

\begin{corollary} Let ${\rm{\bf A}}=[a_{ik}]_{m\times N}\in 
{\Bbb K}^{m\times N}$
$(\Bbb K=\Bbb R$ or  $\Bbb K=\Bbb C)$ be a
matrix with $1\le m\le N$ and $l_2$-normalized columns, i.e.,
  $$
\sum_{i=1}^m|a_{ik}|^2=1\quad{\rm for\,\,  every}\quad k=1,2,\ldots, N.
\leqno(25)
  $$
 Then under notations of Proposition $2.2$, we have
 $$
{\rm Var}[\bar{X}(m)]=\frac{N-m}{m(N-1)}
\left(1-
\frac{1}{N^2}\sum_{i=1}^m|\sum_{k=1}^Na_{ik}|^2\right),\leqno(26)
 $$
or equivalently, 
 $$
\sigma [\bar{X}(m)]:=\sqrt{{\rm Var}[\bar{X}(m)]}=\sqrt{\frac{N-m}{m(N-1)}}
\sqrt{1-\frac{1}{N^2}\sum_{i=1}^m|\sum_{k=1}^Na_{ik}|^2},\leqno(27)
 $$
where $\sigma [\bar{X}(m)]$ is the standard deviation of 
$\bar{X}(m)$.
\end{corollary}

Taking the inequality (12) into the equality  (27) of Corollary 2.5,
we obtain the following estimate.

\begin{corollary} Let ${\rm{\bf A}}=[a_{ik}]_{m\times N}\in 
{\Bbb K}^{m\times N}$
$(\Bbb K=\Bbb R$ or  $\Bbb K=\Bbb C)$ be a
matrix with $1\le m\le N$ and $l_2$-normalized columns, i.e.,
  $$
\sum_{i=1}^m|a_{ik}|^2=1\quad{\rm for\,\,  every}\quad k=1,2,\ldots, N.
  $$
 Then under notations of Proposition $2.2$, we have 
  $$
\sigma [\bar{X}(m)]\le\mu 
\sqrt{1-\frac{1}{N^2}\sum_{i=1}^m|\sum_{k=1}^Na_{ik}|^2}\leqno(28)
 $$
and 
   $$
\sigma [\bar{X}(m)]\le\mu ,\leqno(29)
  $$
where $\sigma [\bar{X}(m)]$ is the standard deviation of 
$\bar{X}(m)$ and $\mu$ is the coherence of the matrix ${\rm{\bf A}}$.
Moreover, the  equality in $(28)$ holds if and only if 
the columns of the matrix ${\rm{\bf  A}}$
form an  equiangular tight frame, while
the  equality in $(29)$ holds if and only if 
the columns of the matrix ${\rm{\bf  A}}$
form an  equiangular tight frame and  the sum of all entries in each of 
these columns of the matrix  ${\rm{\bf A}}$ is equal to zero.  
 \end{corollary}

Finally, we apply Proposition 2.2 to the so-called   
random partial Fourier matrix  
(see, e.g., \cite[p. 372]{fr}, \cite{ct2} and \cite{chj}). Notice that this matrix was 
the first structured random matrix investigated in the theory of compressive sensing 
(see \cite[Section 12.1]{fr}). Namely, this resulting random sensing 
matrix is widely used in compressed sensing, in view of the fact that 
the corresponding memory cost is only $O(N\log N)$ 
(see \cite[Appendix C.1]{fr}). 
  
Recall that the 
{\it disrete Fourier matrix}  ${\rm{\bf F}}\in\Bbb C^{N\times N}$
is a matrix with entries 
  $$
F_{i,k}=\frac{1}{\sqrt{N}}e^{2\pi jik/N},\quad i,k=0,1,\ldots,N-1,\leqno(30)
   $$    
where $j$ is the imaginary unit. It is well known that the Fourier matrix
  ${\rm{\bf F}}$ is unitary. The vector  
${\rm{\bf F}}{\rm{\bf x}}={\rm{\bf \hat{x}}}$ is called the 
{\it disrete Fourier transform} of the vector ${\rm{\bf x}}\in\Bbb C^N$.  
A compressive sensing problem concerns the reconstruction 
a {\it sparse  vector} ${\rm{\bf x}}$ from $m$ independenet 
and uniformly distributed random entries of its discrete Fourier transform
${\rm{\bf \hat{x}}}$. The resulting matrix ${\rm{\bf F}}$ is called 
the {\it random partial Fourier matrix}. Notice that a crucial point 
for applications is that computations can be performed quickly using the 
fast Fourier transform (FFT) (see, e.g., 
\cite[Appendix C.1, pp. 573--575]{fr}) and \cite[Chapter 3]{s1}).

Accordingly,  consider the matrix ${\rm{\bf F}}_m\in \Bbb C^{m\times N}$ 
with $1\le m\le N$ which is formed 
 drawing  independently and uniformly at 
random  $m$ rows of the Fourier matrix  ${\rm{\bf F}}$ defined above. 
Accordingly, if we assume that 
${i_1,i_2,\ldots ,i_m}\subset\{0,1,\ldots, N-1\}$ is a set of indices 
of columns of the matrix  ${\rm{\bf F}}_m$, then taking into account (30),
 its entries are given by 
  $$
F_{i_l,k}=\frac{1}{\sqrt{N}}e^{2\pi ji_lk/N},
\quad l=1,2,\ldots, m; k=0,1,\ldots,N-1.\leqno(31)
   $$
It is known  (see \cite{ct2} and \cite{nt}) that under some conditions on 
$N$, the resulting partial Fourier matrix satisfies the 
{\it restricted isometry property} (introduced in \cite{ct1}) 
with high probability.

Then from Proposition 2.2 easily follows the following result. 

\begin{corollary} Let ${\rm{\bf F}}_m\in \Bbb C^{m\times N}$
be the previously defined matrix with $1\le m\le N$. 
Then under Definition $2.1$ and notations of Proposition $2.2$,
if the matrix ${\rm{\bf F}}_m$ does not contain the first row
of the Fourier matrix ${\rm{\bf F}}\in\Bbb C^{N\times N}$, we have   
  $$
\Bbb E[\bar{X}(m)]=0\leqno(32)
  $$ 
and
 $$
{\rm Var}[\bar{X}(m)]=\frac{N-m}{N(N-1)}.\leqno(33)
 $$
Otherwise we have
$$
\Bbb E[\bar{X}(m)]=\frac{1}{\sqrt{N}}\leqno(34)
  $$ 
and
 $$
{\rm Var}[\bar{X}(m)]=\frac{(N-m)(m-1)}{mN(N-1)}.\leqno(35)
 $$ 

\end{corollary}

\section{Proofs of the results}  

\begin{proof}[Proof of Theorem $1.2$]
By Definition 1.1, we find that
    $$
\Bbb E[X(m,\Phi_N)]=\frac{1}{{N\choose m}}\sum_{\{i_1,i_2,\ldots,i_m\} 
\subset \{1,2,\ldots,N\}}(z_{i_1}+z_{i_2}+\cdots+
z_{i_m}),\leqno(36)
   $$
where the summation ranges over all subsets $\{i_1,i_2,\ldots,i_m\}$
of $\{1,2,\ldots,N\}$ with $1\le i_1<i_2<\cdots <i_m\le N$.
Since any fixed $z_{i_s}$ with $s\in\{1,2,\ldots,N\}$ occurs 
exactly ${N-1\choose m-1}$ times  in the sum on the right hand side 
of (36), and using the identity 
${N\choose m}=\frac{N}{m}{N-1\choose m-1}$, we obtain
     \begin{equation*}\begin{split}
\Bbb E[X(m,\Phi_N)]=&\frac{1}{{N\choose m}}\left({N-1\choose m-1}z_1+
{N-1\choose m-1}z_2+\cdots +{N-1\choose m-1}z_N\right)\\
(37)\qquad\qquad =&
\frac{{N-1\choose m-1}}{{N\choose m}}(z_1+z_2+\cdots +z_N)=
\frac{m}{N}(z_1+z_2+\cdots +z_N),\qquad\qquad
\qquad\qquad\qquad\qquad\qquad\qquad
    \end{split}\end{equation*}
which implies the equality (3).  
 
If $m=1$, then by the additivity of the exceptation of a random variable, 
we have 
  $$
\Bbb E[|X(1,\Phi_N)|^2]=\frac{1}{N}(|z_1|^2+|z_2|^2+\cdots+|z_N|^2).
\leqno(38)
  $$
Notice that (38) coincides with the equality (5) for $m=1$.

Furthermore, by  (3) and (38) we immediately get
  $$
{\rm Var}[X(1,\Phi_N)]  =\Bbb E[|X(1,\Phi_N)|^2]-|\Bbb E[X(1,\Phi_N)]|^2=
\frac{1}{N^2}\left(N  \sum_{i=1}^N|z_i|^2-|\sum_{i=1}^N z_i|^2\right).
\leqno(39)
   $$ 
Note that the expression (39) coincides with the expression (4) for 
$m=1$.

Now suppose that $m\ge 2$. Then  we have 
    $$
\Bbb E[|X(m,\Phi_N)|^2]=\frac{1}{{N\choose m}}\sum_{\{i_1,i_2,\ldots,i_m\} 
\subset \{1,2,\ldots,N\}}(z_{i_1}+z_{i_2}+\cdots+
z_{i_m})(\bar{z_{i_1}+z_{i_2}+\cdots+
z_{i_m}}),\leqno(40)
   $$
where the summation ranges over all subsets $\{i_1,i_2,\ldots,i_m\}$
of $\{1,2,\ldots,N\}$ with $1\le i_1<i_2<\cdots <i_m\le N$.
  Notice that after  multiplication of  terms 
on the right hand side of (40) we obtain that in the obtained sum 
every term of the form $z_i\bar{z}_i= |z_i|^2$  $(i=1,2,\ldots, N)$ 
occurs exactly 
   ${N-1\choose m-1}$ times, while every term of the 
form $z_t\bar{z}_s$ with $1\le t<s\le N$, occurs exactly 
   ${N-2\choose m-2}$ times. Accordingly, the equality (40) becomes  
   $$
\Bbb E[|X(m,\Phi_N)|^2]=\frac{1}{{N\choose m}}\left({N-1\choose m-1}
\sum_{i=1}^N|z_i|^2+
{N-2\choose m-2}\sum_{1\le t<s\le N}z_t\bar{z}_s\right), \leqno(41) 
  $$
whence by using the {\it Pascal's formula} 
${N-1\choose m-1}={N-2\choose m-2}+{N-2\choose m-1}$
 and the identities
${N-2\choose m-1}=\frac{m(N-m)}{N(N-1)}{N\choose m}$ and 
${N-2\choose m-2}=\frac{m(m-1)}{N(N-1)}{N\choose m}$,  
we obtain
   \begin{equation*}\begin{split}
\Bbb E[|X(m,\Phi_N)|^2]=&\frac{1}{{N\choose m}}
\left({N-2\choose m-1}\sum_{i=1}^N|z_i|^2\right.\\
  &+\left.\left({N-2\choose m-2}\sum_{i=1}^N|z_i|^2+
{N-2\choose m-2}\sum_{1\le t<s\le N}z_t\bar{z}_s\right)\right)\\
=&\frac{1}{{N\choose m}}
\left({N-2\choose m-1}\sum_{i=1}^N|z_i|^2
+{N-2\choose m-2}\sum_{1\le t\le s\le N}z_t\bar{z}_s\right)\\
=& \frac{1}{{N\choose m}}\left({N-2\choose m-1}\sum_{i=1}^N|z_i|^2
+{N-2\choose m-2}\left(\sum_{i=1}^N z_i\right)\left(\sum_{s=1}^N 
\bar{z}_i\right)\right)\\
=& \frac{1}{{N\choose m}}\left(\frac{m(N-m)}{N(N-1)}{N\choose m}
 \sum_{i=1}^N|z_i|^2 +\frac{m(m-1)}{N(N-1)}{N\choose m}|\sum_{i=1}^N z_i|^2
\right)\\
=& \frac{m}{N(N-1)}\left((N-m) \sum_{i=1}^N |z_i|^2 + 
 (m-1)\big| \sum_{i=1}^N z_i \big|^2\right).
  \end{split}\end{equation*}
The above equalities yield the expression (5). 
Finally, from the above expression and (3) we have
  \begin{equation*}\begin{split}
{\rm Var}[X(m,\Phi_N)]=& \Bbb E[|X_l(m)|^2]-|\Bbb E[X_l(m)]|^2\\
=& \frac{m}{N(N-1)}\left((N-m) \sum_{i=1}^N |z_i|^2 + 
 (m-1)\big| \sum_{i=1}^N z_i \big|^2\right)-\frac{m^2}{N^2}
|\sum_{i=1}^N z_i|^2\\
=& \frac{m(N-m)}{N^2(N-1)}\left(N \sum_{i=1}^N |z_i|^2
 - \big| \sum_{i=1}^N z_i \big|^2\right).
\end{split}\end{equation*}
  This proves the expression (4) and the proof of Theorem 1.2 is completed.
  \end{proof}

\begin{proof}[Proof of Corollary $1.4$]

Taking the known identity
    $$ 
N\sum_{i=1}^N |z_i|^2-|\sum_{i=1}^Nz_i|^2= \sum_{1\le i<k\le N}|z_i-z_k|^2
   $$
(which can be easily proved by induction on $N$) into the right hand (4)
of Theorem 1.2, immediately gives (6).
  \end{proof}
\begin{proof}[Proof of Proposition $2.2$] 
Using the linearity of the expectation and the expression (3) of
Theorem 1.2, we find that 
   $$
\Bbb E[\bar{X}(m)] =\frac{1}{m}\sum_{i=1}^m\Bbb E[X_i(m)]
=\frac{1}{m}\sum_{i=1}^m\sum_{k=1}^N\frac{m}{N}a_{ik}=
\frac{1}{N}\sum_{i=1}^m\sum_{k=1}^Na_{ik},
  $$
whence it follows the equality (19).

Similarly, since $X_i(m)$  ($i=1,2,\ldots m$) are mutually independent 
random variables (and so, uncorrelated), then using the additivity of the 
variance and the expression (4) of Theorem 1.2, we immediately obtain
   \begin{equation*}\begin{split}
{\rm Var}[\bar{X}(m)]=\frac{1}{m^2}\sum_{i=1}^m{\rm Var}[X_i(m)]=
\frac{N-m}{mN^2(N-1)}
\left(N \sum_{i=1}^m\sum_{k=1}^N|a_{ik}|^2-
\sum_{i=1}^m|\sum_{k=1}^Na_{ik}|^2\right),  
  \end{split}\end{equation*}
which is actually the expression (20).

Finally, taking the expression for $\Vert{\rm{\bf A}}\Vert_F$ given by (14)
 into (20), immediately gives the equality (21), which completes the proof 
of Proposition 2.2.
  \end{proof}

\end{document}